\documentclass[journal=ancac3, manuscript=article]{achemso}

\usepackage{chemformula} 

\usepackage{graphicx,xcolor}
\usepackage{amsmath}

\author{Krishnakanta Mondal}
\affiliation{Department of Physics, Indian Institute of Science Education and Research, Dr. Homi Bhabha Road, Pune-411008, India}
\altaffiliation{Current affiliation: Department of Physical Sciences, Central University of Punjab, Bathinda, Punjab-151001, India}
\author{Prasenjit Ghosh}
 \email{prasenjit.jnc@gmail.com} 
\affiliation{Department of Physics, Centre for Energy Sciences, 
              Indian Institute of Science Education and Research, Dr. Homi Bhabha Road, Pune-411008, India
            }

\title{Exfoliation of Ti$_2$C  and Ti$_3$C$_2$ Mxenes from Bulk Phases of Titanium Carbide: A Theoretical Prediction}


\begin{document}

\begin{abstract}
MXenes, a new class of two dimensional materials with several novel properties, are usually prepared from their MAX phases by etching out the A element using strong chemical reagents. This results in passivation of the surfaces
of MXene with different functional groups like O, -OH, -F, etc., which in many cases tend to degrade their properties.
In this work, using first principle density functional theory based calculations, we propose a novel method to synthesize pristine Ti$_2$C and Ti$_3$C$_2$ MXenes
from the bulk titanium carbides with corresponding stoichiometry. Based on the values of cleavage energy obtained from our calculations,
we envisage that pristine Ti$_2$C and Ti$_3$C$_2$ MXenes can be prepared, using mechanical or sonification-assisted liquid-phase exfoliation techniques,
from their bulk phases.
\end{abstract}

\section{Introduction}
MXenes, with a general formula of M$_{n+1}$X$_n$ (M represents a transition metal element, while X can be carbon or nitrogen), are the latest additions to the family of
two-dimensional (2D) materials\cite{naguib11,naguib13,naguib12,khazaei13,gogotsi15,anasori17}. These newly discovered materials are being
extensively investigated due to their novel structural, chemical, optical, and magnetic properties \cite{tang12,xie14,naguib13,lukatskaya13,naguib12,bai16,guo16},
which make them promising candidates for various applications such as catalysis, anode materials for Li-ion battery, hydrogen storage, supercapacitors, environmental pollutant decontaminators \cite{tang12,xie14,naguib13,lukatskaya13,naguib12}, etc. Till date 19 MXenes are experimentally realized
and about 50 have been theoretically predicted to be stable\cite{anasori17}.

These MXenes are mainly synthesized from the MAX phase by chemical etching of A-metal, where A is a Group 13 or 14 element \cite{naguib14}.
Typically HF, HCl, LiF and NH$_4$HF are used as etching reagents\cite{anasori17,naguib14,ghidiu14,wang16,halim14,karlsson15}.
The surfaces of MXenes prepared from the MAX phases using this method are usually passivated with various chemical groups like
-OH, -O, -F, -H etc. Often this surface passivation has degrading effects on the performance of the MXenes.
For example, Tang et al., theoretically predicted that the Li ion storage capacity of Ti$_3$C$_2$ MXene is larger than that
of the -OH or F passivated ones\cite{tang12}. Moreover, Junkaew \textit{et al.} reported that the chemical activity of the pristine
Ti$_2$C is higher than that of the O-terminated one\cite{junkaew18}. Further, the magnetism of Ti$_2$C is quenched
due the termination of surfaces with O-atom or -OH group\cite{xie13}. Additionally, due to the unavailability of the pristine MXenes,
the experimental researchers are not able to verify several of the theoretically predicted superior properties of MXenes.
Therefore, it is highly desirable to develop novel methods, that do not use etching agents, for synthesis of high quality pristine MXenes.

In view of preparing pristine MXene, in 2015, a bottom up approach, with chemical vapour deposition (CVD) method, was used to synthesize 2D layers of $\alpha$-Mo$_2$C\cite{gogotsi15}. Using this method a ultrathin (few nano meter) layer of $\alpha$-Mo$_2$C was prepared. Similar method was also used to synthesize ultrathin layers of tungsten and tantalum carbides\cite{xu15}. However, the synthesis of a monolayer of MXene is yet to be demonstrated\cite{anasori17}. A method leading to the realizaton of pristine MXene will be a breakthrough in this field. To this end, in this work, using first principles density functional theory (DFT) based calculations,
we have proposed a novel process that uses physical methods to synthesize Ti$_2$C and Ti$_3$C$_2$ MXenes from
their corresponding titanium carbide bulk phases \textit{without} using etching agents.

\section{Computational details}
All the calculations were performed with the Quantum ESPRESSO (QE) package \cite{qe09,qe17} which is an implementation
of DFT in a plane wave pseudopotential framework. 
The electron-electron exchange and correlation functional was
described with the Perdew-Burke-Ernzerhof (PBE)\cite{pbe} parametrization of the generalized gradient approximation(GGA). To include the van der Waals correction we
have used Grimme's dispersion method as implemented in QE\cite{grimme06}.
The calculations were performed employing kinetic energy cutoffs of 55 and 480 Ry for the wave 
function and augmentation charge density, respectively. To speed up the convergence, we have used the Marzari-Vanderbilt smearing with a smearing width of 0.007 Ry\cite{mv99}. We have modelled the surfaces of Ti$_2$C and Ti$_3$C$_2$ using six and four layered slabs respectively. Further, to avoid the interactions
between the periodic images we have used a vacuum of more than 16 \AA\, along z-axis. We have ensured that this vacuum is maintained even for the case
where we have displaced the top most layer to a distance of 10 \AA~ from the layer below it to mimic the exfoliation process.
The Brillouin zone has been sampled using 12$\times$ 12 $\times$ 1 k-points for the slab calculation.

To analyse the interlayer interaction in the bulk phases of Ti$_2$C and Ti$_3$C$_2$ we have carried out NCI (Non Covalent Interaction)\cite{johnson10, julia11} analysis, as implemented in CRITIC2 code\cite{critic2, alberto12}, on these systems. NCI analysis is based on the electron density, $\rho$ and its reduced density gradient, s, which is defied in the following,
\begin{equation}
    s = \frac{1}{2(3\pi^{2})^{1/3}}  \, \frac{\left|\nabla \rho\right|}{\rho^{4/3}}
\end{equation}
The isosurface of s in 3D have been visualised using VMD. The isosurface is colored (blue-green-red color scale) according to the value of
 sign($\lambda_2$)$\rho$. 
 Blue indicates the weak attractive interaction and green represents the van der Waals interaction while red shows non-bonding interaction.

The formation energy ($E_{form}$) of the bulk phases of Ti$_2$C has been calculated using the following equation:
\begin{equation}
E_{form} = \frac{E(Ti_mC_n) - m\times\mu(Ti_{hcp}) - n\times\mu(C_{grap})}{n+m}
\end{equation}
where $E(Ti_mC_n)$ is the energy of the bulk unitcell of Ti$_2$C and  $\mu(Ti_{hcp})$ and $\mu(C_{grap})$ are the chemical potentials of Ti and C atoms, respectively. $\mu(Ti_{hcp})$ has been calculated from the bulk HCP phase of Ti and $\mu(C_{grap})$ is taken from the graphite phase of C. $m$ and $n$ indicate the number of Ti and C atoms in the bulk phase of Ti$_2$C, respectively.

\section{Results and Discussion}

\begin{figure}[h]
\centering
\includegraphics[scale=.43]{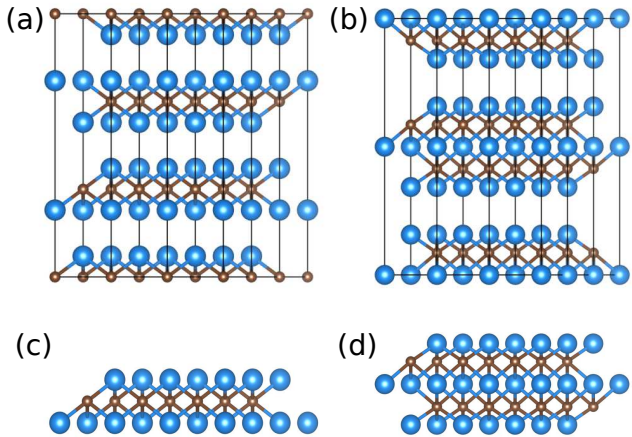}
\caption{Side view of the structure of bulk phase and MXene of Ti$_2$C (a, c) and Ti$_3$C$_2$ (b, d) . In this and the subsequent figures in this manuscript
the Ti and C atoms are represented by large blue spheres and small brown spheres, respectively.} 
\label{fig:bulk-ti2c}
\end{figure}

We begin by noting that amongst the many polytypes of bulk Ti$_2$C, the two most stable ones are the trigonal (Figure \ref{fig:bulk-ti2c}(a)) and
the cubic (Figure S1(a)) phases \cite{em96,eibler07,frisk03,hugosson01,tashmetov02}. Depending on the experimental conditions, both
these phases have been successfully synthesized\cite{em96,frisk03,tashmetov02}. From our calculations we find that, in agreement with previous
literature report\cite{eibler07}, the cubic phase is about 0.02 eV/atom more stable than the trigonal phase (Table S1). While the
cubic phase of Ti$_2$C have a sodium chloride structure with the C atoms or vacancy surrounded by the distorted octahedra of Ti atoms,
the trigonal phase forms a layered structure of the form Ti-C-Ti, where the Ti atoms in the two hexagonal planes
are separated by a plane of C atoms. We note that each Ti-C-Ti layer of the bulk trigonal phase is similar to that of a layer of Ti$_2$C MXene (Figure \ref{fig:bulk-ti2c}(c)).
Moreover, the in plane lattice parameter of 3.08 \AA~of the trigonal phase is also similar to that of the Ti$_2$C MXene lattice parameter of 3.01 \AA\,\cite{kurtoglu12}.

Analogous to the layered trigonal structure of bulk Ti$_2$C, bulk Ti$_3$C$_2$ also have a layered structure with hexagonal symmetry, the layers being stacked along the (0001) 
direction (Figure \ref{fig:bulk-ti2c}(b)).
Each layer in the bulk hexagonal phase of Ti$_3$C$_2$ is constructed with the atomic arrangement of Ti-C-Ti-C-Ti, where the C atoms are sandwiched between Ti layers. It is observed 
that the in plane lattice parameters (3.07 \AA) of the hexagonal phase of Ti$_3$C$_2$ is similar to those of the corresponding MXene (3.10 \AA)\cite{bai16}. Therefore, each layer of Ti-C-Ti-C-Ti in the bulk hexagonal phase of Ti$_3$C$_2$ can be identified as the well known Ti$_3$C$_2$ MXene.

The above observations of the exceptional similarity between the layered structures of the trigonal (hexagonal) phase of bulk Ti$_2$C (Ti$_3$C$_2$) with that of
Ti$_2$C (Ti$_3$C$_2$) MXene stimulated us to ask the question: Is it possible to mechanically exfoliate a single layer of Ti$_2$C (Ti$_3$C$_2$) from the bulk trigonal (hexagonal) phase?
If this can be achieved, then it will provide a novel route to synthesize highly desirable pristine Ti$_2$C and Ti$_3$C$_2$ MXenes.

\begin{figure}
\centering
\includegraphics[scale=0.50]{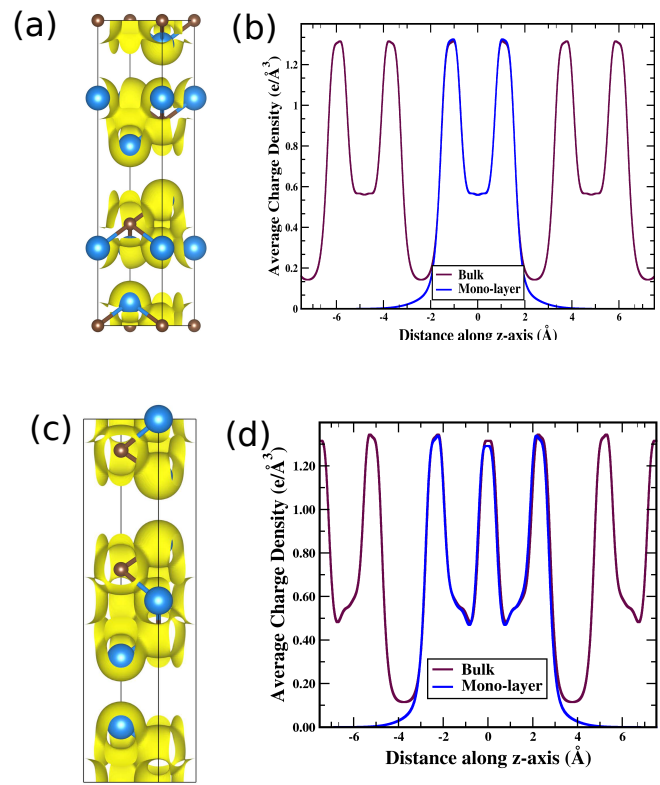}
\caption{Charge density isosurfaces of the bulk (a) Ti$_2$C, and (c) Ti$_3$C$_2$. Planar average of charge density of MXene (b) Ti$_2$C, and (d) Ti$_3$C$_2$.} 
\label{fig:charge}
\end{figure}

To provide an answer to the above mentioned question it is imperative to understand the nature and strength of the interaction between two MXene layers in the bulk phases of Ti$_2$C and Ti$_3$C$_2$. 
The charge density distribution between two such layers in the bulk trigonal (hexagonal) phase will provide an indication of the nature and strength of interactions between
them. A strong
covalent interaction between two layers will result in significant accumulation of charge density between them while weak metallic/van der Waals
interaction will result in negligible charge density between two such layers. Figure \ref{fig:charge}(a) (Figure \ref{fig:charge}(c)) shows the charge density isosurface plot for the layered bulk
phase of Ti$_2$C (Ti$_3$C$_2$). To quantify the interlayer charge density further, we have also plotted in Figure \ref{fig:charge}(b) (Figure \ref{fig:charge}(d)) the planar average (averaged over the $xy$-plane) of the charge
density as a function of $z$ along the (0001) direction in bulk Ti$_2$C (Ti$_3$C$_2$). From Figure \ref{fig:charge} we find that for both Ti$_2$C and Ti$_3$C$_2$ the charge is localized within the layer with negligible
charge density in between the two layers. Further for comparison with Ti$_2$C (Ti$_3$C$_2$) MXene, we have also plotted the planar average of the charge density of MXene in
Figure \ref{fig:charge}(b) (Figure \ref{fig:charge}(d)). We find that the two charge density profiles are almost identical. 

\begin{figure}[h]
\centering
\includegraphics[scale=.25]{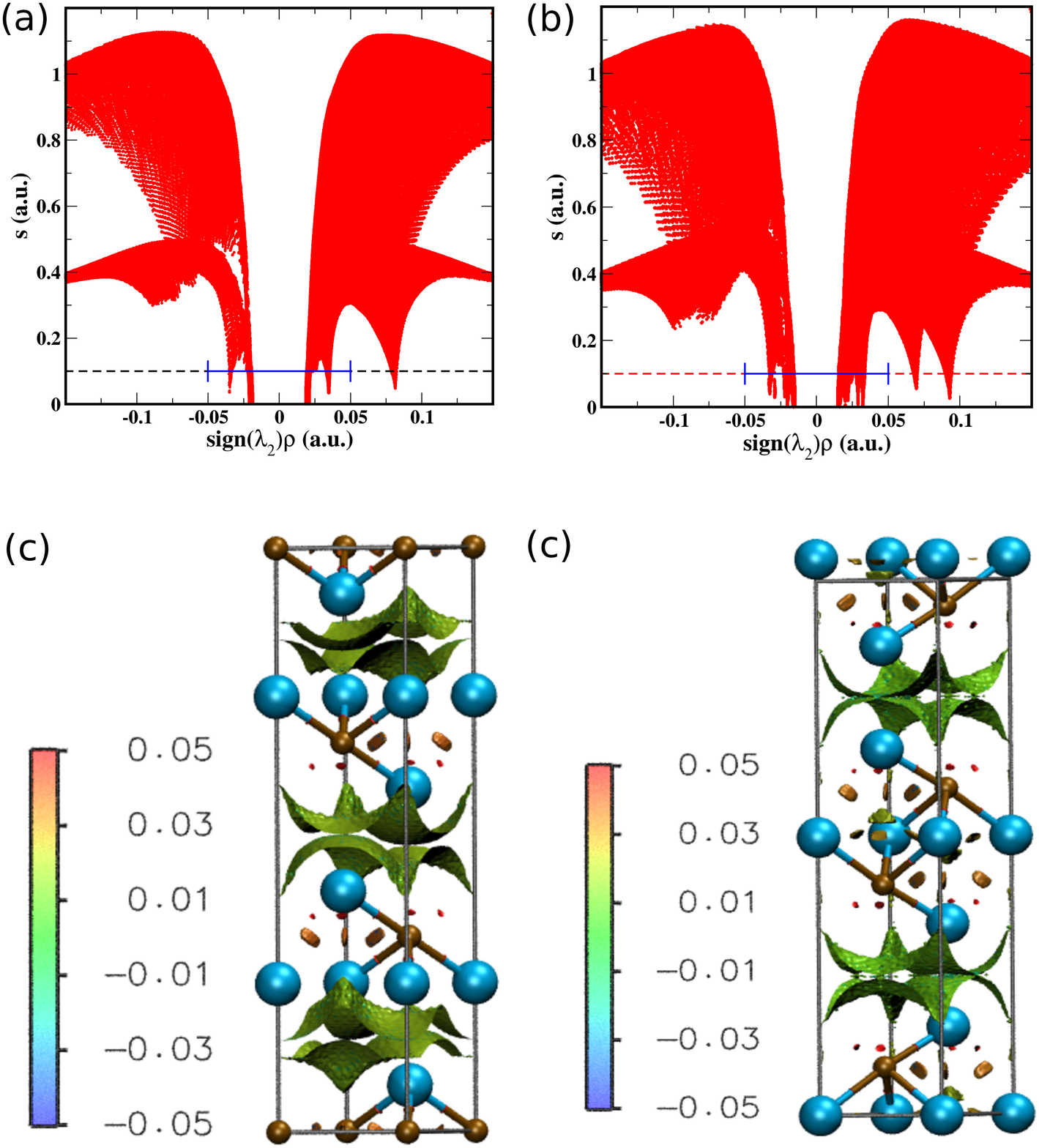}
\caption{NCI plots for the bulk phases of (a,c) Ti$_2$C and (b,d) Ti$_3$C$_2$. NCI plots have been shown for s=0.1 a.u. (dotted line in (a) and (b)). The color scale shows the range of sign($\lambda_2$)$\rho$ between -0.05 and 0.05 a.u. } 
\label{fig:nciplot}
\end{figure}

Furthermore, to get more insight into the bonding be-
tween the layers we have carried out NCI (Non Covalent
Interaction)\cite{johnson10, julia11} analysis on the bulk phase of Ti$_{2}$C (Ti$_3$C$_2$). Figure \ref{fig:nciplot}(a) and (b) show the s vs. sign($\lambda_2$)$\rho$ for Ti$_2$C and Ti$_3$C$_2$ respectively. We find that there are tails of s at very low negative values of sign($\lambda_2$)$\rho$) suggesting that there might be weak attractive interactions between the
layers. Further we plotted the isosurface of s for s =0.1 (denoted by the blue horizontal line in Figure \ref{fig:nciplot}(a) and (b)) for sign($\lambda_2$)$\rho$ between -0.05 and 0.05 a.u. These isosurfaces
are shown in Figure 3(c) and (d). The plots show greenish isosurfaces at the interlayer positions thereby further supporting our claim that the interaction between two such
layers are indeed weak and of non-covalent origin.

From the above discussion it is clear that the
interlayer interaction in bulk Ti$_2$C (Ti$_3$C$_2$) is significantly weak which suggest that a single layer can be exfoliated
from a slab of Ti$_2$C (Ti$_3$C$_2$).

\begin{figure}[h]
\centering
\includegraphics[scale=0.530]{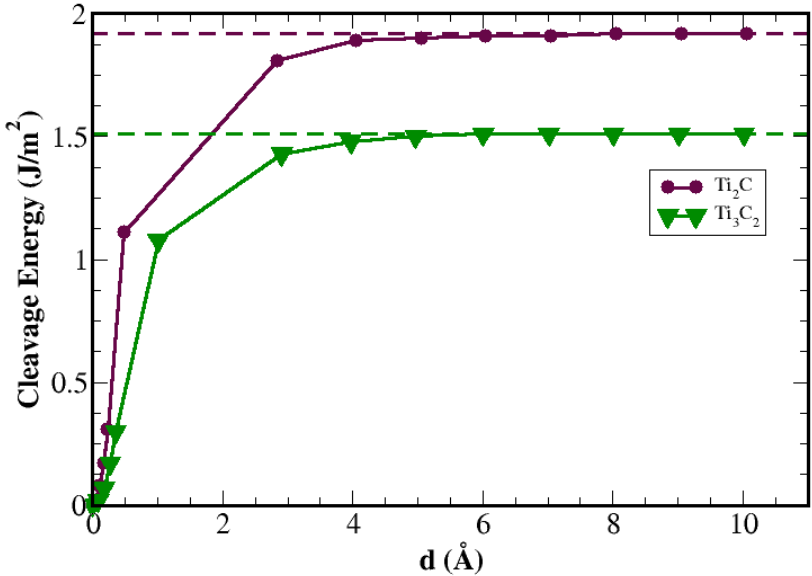}
\caption{Cleavage energy of the MXenes as a function of the distance ($d$) of the cleaved layer from the surface.} 
\label{fig:exfoil}
\end{figure}

Since we have ascertained that the interlayer interactions in bulk trigonal Ti$_2$C and hexagonal Ti$_3$C$_2$ is weak, we proceed to compute the energy
cost of exfoliating a single layer of Ti-C-Ti or Ti-C-Ti-C-Ti from their corresponding bulk phases. 
First, we have investigated the exfoliation of Ti-C-Ti layer from the trigonal phase of Ti$_2$C. For this purpose we have chosen the (0001) surface of the bulk phase,
which is represented by a slab with a thickness of six layers (Slab6) (see Figure S3). From this slab, we gradually
exfoliate the surface Ti-C-Ti layer by slowly increasing interlayer distance between the surface Ti-C-Ti layer and that
below it. At each step we have performed a constrained relaxation keeping the z-coordinate of the C-atom of the exfoliated surface layer and the middle C-atom of the
slab fixed. This step-wise detachment of the surface layer is done till the exfoliated layer does not interact with the rest of the slab.
The cleavage energy ($E_{cl}$) has been calculated using the following equation:
\begin{equation}
E_{cl} = \frac{1}{2A} [E(slab6) - E^d(slab6)]
\label{eq:exfoil}
\end{equation}
\noindent where $E(slab6)$ is the energy of the Slab6 and E$^d(slab6)$ indicates the energy corresponding to the Slab6 when the topmost layer is shifted to a distance $d$
away from the Slab6. $E_{cl}$ as a function of $d$ is plotted in Figure \ref{fig:exfoil}. $A$ denotes the area of the surface unit cell.
We find that the cleavage energy for the single layer is 1.82 J/m$^2$. Moreover, to check whether it is easier to cleave a single Ti-C-Ti layer or two such
layers during exfoliation, we have also computed the cleavage energy for the exfoliation of bilayer of Ti$_2$C MXene. For the bilayer, we determine
$E_{cl}$ to be 1.78 J/m$^2$ which is slightly less than that of the single layer. This suggests that during the exfoliation process both bilayers and monolayers
will be present. We note that from two such bilayers, single layers can also be exfoliated.

Following similar exfoliation process described above we have calculated the cleavage energy for the exfoliation of Ti$_3$C$_2$ layer from its bulk hexagonal phase. 
 In Figure \ref{fig:exfoil} we have plotted $E_{cl}$ \textit{vs} $d$ for the case of Ti$_3$C$_2$. The details of the exfoliation process used in the calculations
 have been described in the Supporting Information. 
We found the cleavage energy for the exfoliation of Ti$_3$C$_2$ to be 1.51 J/m$^2$, which is 0.31 J/m$^2$ lower than the corresponding value for the case of Ti$_2$C. This indicates that the exfoliation of Ti$_3$C$_2$ MXene from its bulk phase is easier than that for the case of Ti$_2$C MXene. 

In order to provide a perspective as to how easy or difficult to exfoliate a single layer of Ti$_2$C or Ti$_3$C$_2$, we compare our computed cleavage energy with those
reported for other layered materials in literature. For example for exfoliating a single atomic layer thick graphene sheet from bulk graphite the cleavage energy is about
0.37 J/m$^2$ \cite{jing17}. This tells us that it is about 4-5 times easier to exfoliate graphene from graphite compared to our systems.
In contrast, for exfoliating quasi-two dimensional (more than one atomic layer thick) sheets, for example GeP$_3$ layer from its bulk, the cleavage energy is 1.14 J/m$^2$, which is about 0.68 J/m$^2$ (0.37 J/m$^2$) higher than that we have observed for Ti$_2$C (Ti$_3$C$_2$).
Although the cleavage energy for exfoliating Ti$_2$C (Ti$_3$C$_2$) is larger than those observed for exfoliation of 2D layers from different
van der Waals solids, we would like to mention that very recently layered 2D materials have been exfoliated from non-van der Waals
solids also. For example, Balan \textit{et al.} have synthesized a novel 2D material ``hematene" from natural iron-ore hematite ($\alpha$-Fe$_2$O$_3$)
using liquid exfoliation technique\cite{balan18}. Using a sonification-assisted liquid-phase exfoliation method Yadav \textit{et al.}
demonstrated that it is possible to synthesize magnetic 2D material ``chromiteen" from naturally occurring mineral chromite (FeCr$_2$O$_4$)\cite{yadav18}.
For both these cases, the exfoliation results in cleavage of strong covalent bonds. This suggests that using similar experimental
techniques, it is possible to exfoliate Ti$_2$C and Ti$_3$C$_2$ MXenes from their corresponding layered bulk phases. We note that in this case the exfoliation
would involve breaking of weaker metallic bonds between the Ti atoms of two interacting Ti-C-Ti or Ti-C-Ti-C-Ti layers depending on the systems.

\section{Conclusions}

In summary, we have proposed a novel method to synthesize pristine Ti$_2$C and  Ti$_3$C$_2$ MXenes by exfoliation of, respectively, Ti-C-Ti and Ti-C-Ti-C-Ti layers from their corresponding layered bulk phases. Based on our computed cleavage energy and some recent experimental reports on synthesis of layered magnetic
2D materials from non-van der Waals solids using sonification-assisted liquid-phase exfoliation method, we suggest that use of
similar methods in this case will enable experimentalists to exfoliate highly desired pristine Ti$_2$C and Ti$_3$C$_2$ MXenes. We hope that our results will motivate experimentalists
to use our proposed methodology to try synthesizing pristine Ti$_2$C and Ti$_3$C$_2$ MXenes.


\section*{Acknowledgments}

PG would like to acknowledge Dr. Nirmalya Ballav, IISER Pune for helpful discussions. KM and PG would like to acknowledge Department
of Scinece and Technology, India Grant No: EMR/2016/005275 for funding. PG would like to acknowledge Department of Science and Technology-Nanomission, India Grants No: SR/NM/NS-15/2011, SR/NM/NS-1285/2014 and SR/NM/TP-13/2016
for funding.

\providecommand{\latin}[1]{#1}
\providecommand*\mcitethebibliography{\thebibliography}
\csname @ifundefined\endcsname{endmcitethebibliography}
  {\let\endmcitethebibliography\endthebibliography}{}


\pagebreak

\setcounter{figure}{0}
\makeatletter 
\renewcommand{\thefigure}{S\@arabic\c@figure}
\makeatother

\section*{Suplementary of Exfoliation of Ti$_2$C  and Ti$_3$C$_2$ Mxenes from Bulk Phases of Titanium Carbide: A Theoretical Prediction}

\begin{figure}[h]
\centering
\includegraphics[scale=4]{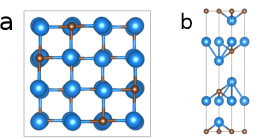}
\caption{Different stable phases of bulk Ti$_2$C, (a) Cubic, and (b) Trigonal} 
\end{figure}

\begin{table}[h]
\centering
\caption{Comparison of our computed results with those of the reported values \cite{eibler07,frisk03,hugosson01}. The values within the parenthesis indicate the experimental data taken from ref. \cite{frisk03}. The results obtained with the van der Waals correction are given within the square brackets.}
\begin{tabular}{ccccc}
\hline
		& \multicolumn{2}{c}{Cubic} &	\multicolumn{2}{c}{Trigonal}			\\
\hline
	 &	our results	&	reported	&	our results	&	reported 	\\
\hline
E$_{form}$	& -0.66 [-0.71]		&	-0.68 [-0.70]	&	-0.64	&	-	\\
a(\AA)	&	8.64 [8.62]	&	8.64 (8.60)	&	3.08 [3.09]	&	3.09 (3.06) 	\\
c(\AA)	&	-	&	-	&	14.45 [14.48]	&	14.45 (14.91) 	\\                                  
\hline
\end{tabular}
\label{tab:comp}
\end{table}

\section*{Slab Convergence Test}
\begin{figure}[h]
\centering
\includegraphics[scale=.54]{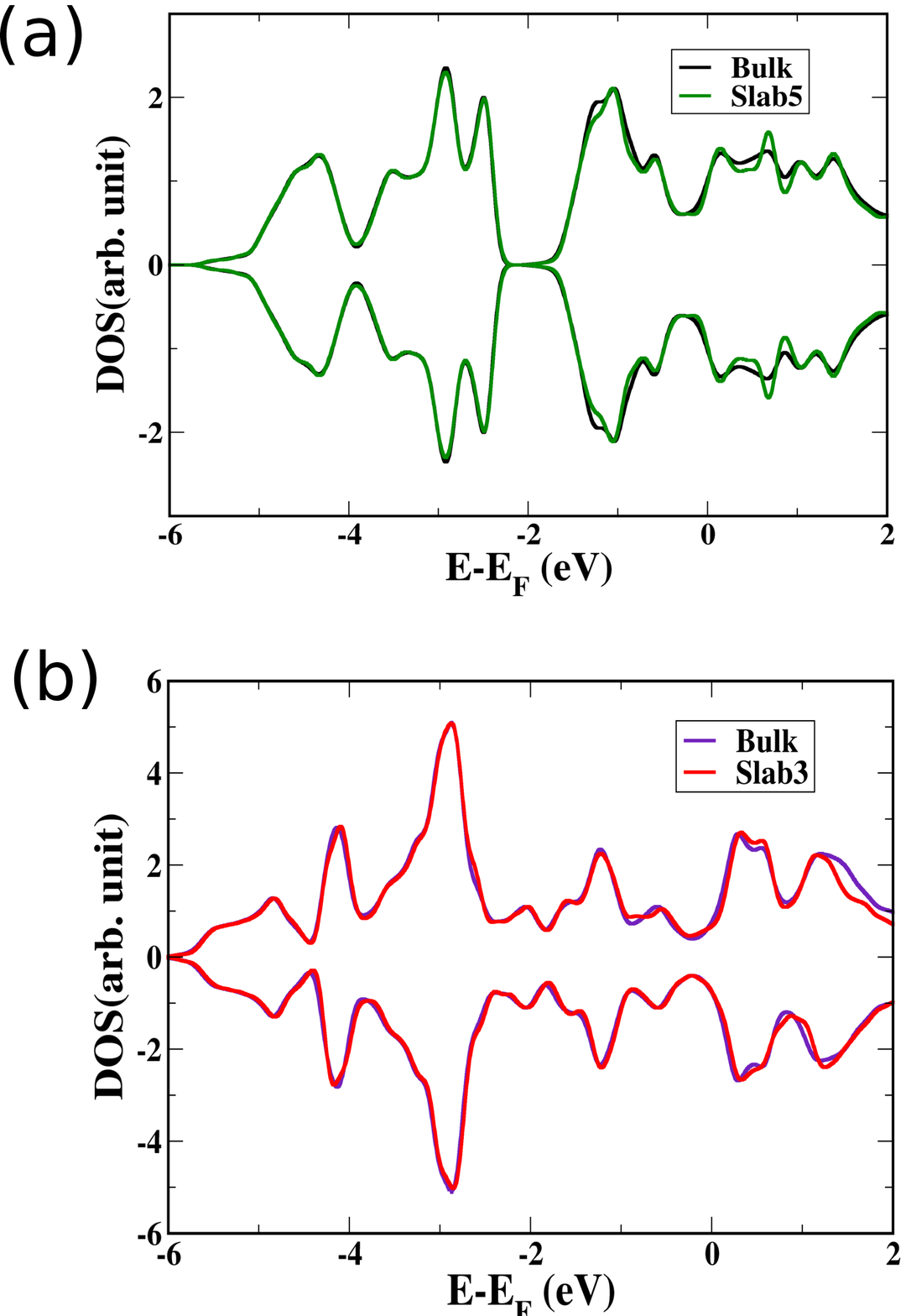}
\caption{Density of states of (a) Ti$_2$C and (b) Ti$_3$C$_2$.} 
\label{fig:dos-ti2c-ti3c2-slab}
\end{figure}

It is found that the five layers of Ti$_2$C, which we named as Slab5, is required to achieve the convergence of the thickness of the slab along (0001). In Figure \ref{fig:dos-ti2c-ti3c2-slab} we have plotted the density of the states (DOS) of the middle layer of this slab along with that of the bulk
From the Figure \ref{fig:dos-ti2c-ti3c2-slab}(a) it can be observed that the density of the bulk closely matches with that of the middle layer in slab5.
This clearly indicates that the Slab5 can  mimic a piece of bulk Ti$_2$C.
For the case of Ti$_3$C$_2$, it is found that the three layers of Ti$_3$C$_2$, which we named as Slab3, is required to achieve the convergence of the thickness of the slab along (0001). In Figure \ref{fig:dos-ti2c-ti3c2-slab}(b) we have plotted the density of the states (DOS) of the middle layer of this slab along with that of the bulk
From the Figure \ref{fig:dos-ti2c-ti3c2-slab} (b) it can be observed that the density of the bulk closely matches with that of the middle layer in slab3.
This clearly indicates that the Slab3 can  mimic a piece of bulk Ti$_3$C$_2$.

\section*{Exfoliation of T$\textsc{i}_3$C$_2$ MXene}

We have investigated the exfoliation of Ti-C-Ti-C-Ti layer from the trigonal phase of Ti$_3$C$_2$. 
\begin{figure}[h]
\centering
\includegraphics[scale=.4]{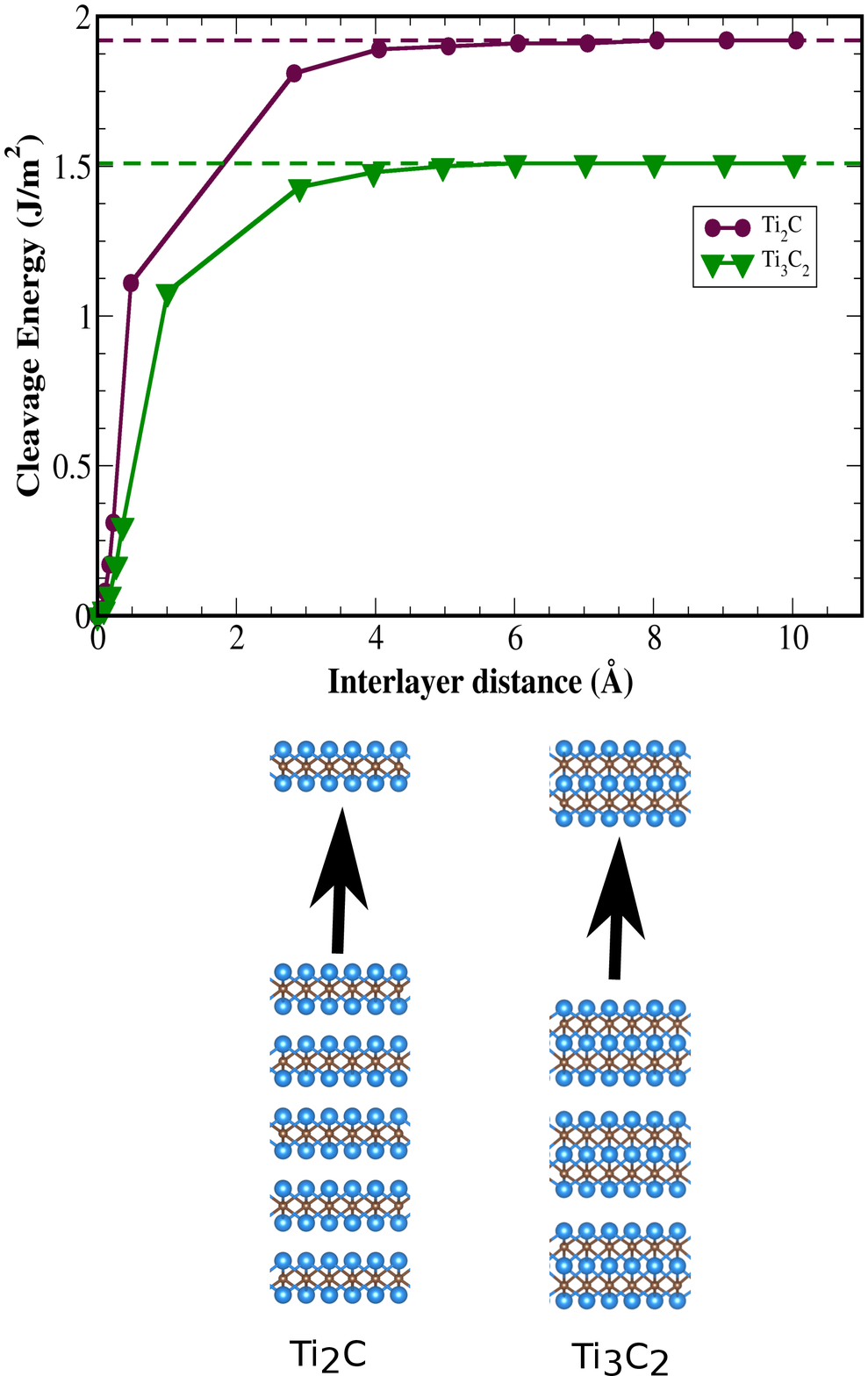}
\caption{Exfoliation of Ti$_2$C and Ti$_3$C$_2$.} 
\label{fig:exfoil-ti2c-ti3c2}
\end{figure}
For this purpose we have chosen the (0001) surface of the bulk phase,
which is represented by a slab with a thickness of 4 layers (Slab4) (see Figure \ref{fig:exfoil-ti2c-ti3c2}). From this slab, we gradually
exfoliate the surface Ti-C-Ti-C-Ti layer.
This is done by slowly increasing interlayer distance between the surface Ti-C-Ti-C-Ti layer and that
below it. At each step we have performed a constrained relaxation keeping the z-coordinate of the middle Ti-atom of the exfoliated surface layer and the middle Ti-atom of the
slab fixed. This step-wise detachment of the surface layer is done till the exfoliated layer does not interact with the rest of the slab.

\end{document}